# One-step exfoliation method for plasmonic activation of large-area 2D crystals

Qiang Fu†, Jia-Qi Dai†, Xin-Yu Huang†, Yun-Yun Dai, Yu-Hao Pan, Long-Long Yang, Zhen-Yu Sun, Tai-Min Miao, Meng-Fan Zhou, Lin Zhao, Wei-Jie Zhao, Xu Han, Jun-Peng Lu, Hong-Jun Gao, Xing-Jiang Zhou, Ye-Liang Wang*, Zhen-Hua Ni*, Wei Ji* and Yuan Huang*

Affiliations

**Advanced Research Institute of Multidisciplinary Science, Beijing Institute of Technology, Beijing 100081, P. R. China**

Qiang Fu, Xin-Yu Huang, Yun-Yun Dai, Xu Han, Ye-Liang Wang, and Yuan Huang

**School of Physics and Key Lab of MEMS of the Ministry of Education, Southeast University, Nanjing 211189, P. R. China**

Qiang Fu, Meng-Fan Zhou, Wei-Jie Zhao, Jun-Peng Lu, and Zhen-Hua Ni

**Institute of Physics, Chinese Academy of Science, Beijing 100190, P. R. China**

Qiang Fu, Long-Long Yang, Zhen-Yu Sun, Xu Han, Tai-Min Miao, Hong-Jun Gao, Xing-Jiang Zhou, and Yuan Huang

**Department of Physics and Beijing Key Laboratory of Optoelectronic Functional Materials & Micro-Nano Devices, Renmin University of China, 100872, Beijing, P. R. China**

Jia-Qi Dai and Wei Ji

**China North Vehicle Research Institute, Beijing 100072, P. R. China**

Yu-Hao Pan

**Songshan Lake Materials Laboratory, 523808, Dongguan, P. R. China**

Lin Zhao and Xing-Jiang Zhou

**University of Chinese Academy of Sciences, 100049, Beijing, P. R. China**

Hong-Jun Gao, Xing-Jiang Zhou

†These authors contributed equally to this work.

Correspondence to: yeliang.wang@bit.edu.cn (Y.L.W.), zhni@seu.edu.cn (Z.H.N.), wji@ruc.edu.cn (W.J.), and yhuang@bit.edu.cn (Y.H.)

## Abstract

Advanced exfoliation techniques are crucial for exploring the intrinsic properties and applications of 2D materials. Though the recently discovered Au-enhanced exfoliation technique provides an effective strategy for preparation of large-scale 2D crystals, the high cost of gold hinders this method from being widely adopted in industrial applications. In addition, direct Au contact could significantly quench photoluminescence (PL) emission in 2D semiconductors. It is therefore crucial to find alternative metals that can replace gold to achieve efficient exfoliation of 2D materials. Here, we present a one-step Ag-assisted method that can efficiently exfoliate many large-area 2D monolayers, where the yield ratio is comparable to Au-enhanced exfoliation method. Differing from Au film, however, the surface roughness of as-prepared Ag films on $SiO_2$/Si substrate is much higher, which facilitates the generation of surface plasmons resulting from the nanostructures formed on the rough Ag surface. More interestingly, the strong coupling between 2D semiconductor crystals (*e.g.* $MoS_2$, $MoSe_2$) and Ag film leads to a unique PL enhancement that has not been observed in other mechanical exfoliation techniques, which can be mainly attributed to enhanced light-matter interaction as a result of extended propagation of surface plasmonic polariton (SPP). Our work provides a lower-cost and universal Ag-assisted exfoliation method, while at the same offering enhanced SPP-matter interactions.

## Introduction

2D materials and their heterostructures offer a unique platform for the exploration of different physical phenomena at the atomic-scale limit, including quantum Hall effects,[1, 2] moiré heterostructures related physics,[3, 4] superconductivity[5], charge density waves (CDW)[6] and magnetism[7]. These intriguing properties are enabling diverse device applications such as field effect transistors (FETs),[8, 9] quantum emitters,[10-12] memory devices,[13] and energy storage units.[14] The continuous improvement of 2D material fabrication methods plays a crucial role in enabling the discovery of novel properties and device applications of 2D materials. Chemical vapor deposition (CVD) and mechanical exfoliation are two most-commonly used methods for obtaining mono- and few-layers crystals. But despite CVD methods showing a clear advantage for preparing wafer-scale monolayers,[15, 16] or even twisted heterobilayers[17], the in-plane strain and higher density of defects and impurities in CVD-grown samples are still challenging to control.[12, 18, 19]

As one of the most widely used "top-down" 2D materials fabrication strategy, mechanical exfoliation shows advantages in feasibility and cost-effectiveness for exploring their novel properties. Traditional exfoliation method via scotch tape or polydimethylsiloxane (PDMS) tape can be employed for getting high-quality 2D monolayer crystals, but are limited in terms of sample size (~10-100 μm) and exfoliation yield. An oxygen plasma cleaning method[20] was proposed to enhance the yield and size of graphene and layered copper oxide high temperature superconductors. And more recently, Au-enhanced exfoliation[21-24] offers a universal and one-step approach to prepare large-area 2D crystals. Although Au-enhanced exfoliation shows obvious advantages for exploring many intrinsic physical properties of 2D crystals, for some optical measurements, especially PL spectroscopy, this method is severely hindered because of the quenching effect.[21, 23-25] Au is a material also too costly for massive industrial production of 2D crystals. Another metal in IB group, silver (Ag), is significantly cheaper than gold, and this study explores in detail the feasibility of using Ag to replace Au to exfoliate 2D crystals, and other potential advantages beyond just cost saving.

In this work, to test the feasibility of Ag-assisted exfoliation, we first perform theoretical calculations on the adhesive energy of different layered materials with silver. From there, we introduce a contamination-free, one-step and universal Ag-assisted mechanical exfoliation method, which has been successfully used to exfoliate 12 types of single-crystalline monolayers with millimeter-size, including metal-dichalcogenides, black phosphorus (BP), 2D magnets and superconductors. The exfoliation efficiency of this Ag-assisted method is almost identical to that of the Au-enhanced exfoliation method. Interestingly, we also observed that the activated plasmon on the surface of Ag film is enhancing the PL emission of some exfoliated 2D semiconductors, like $MoS_2$ and $MoSe_2$, which is significantly higher when compared to the PL intensity generated from 2D semiconductors exfoliated on Au film. Our results demonstrate that the Ag-assisted exfoliation method can provide a robust and lower cost approach to prepare emergent 2D materials, along with the capability to generate strong SPP field enhancement at the metal – 2D material interface.

**Results**

Prediction of Ag-assisted exfoliation

Density functional theory (DFT) calculations were carried out to examine whether Ag thin film is a good candidate for assisted exfoliation, looking at specifically the relative adhesive energies and perturbation to the electronic structures of 2D layers. We used the ratio $R_{MA/IL}$ to represent the layer-Ag over the layer-layer adhesive energies[23]. Fig. 1a and Supplementary Table 1 show the comparison of 18 representative 2D materials, which show an $R_{MA/IL}$ range of 1.35 to 2.92 for 16 types of 2D crystals. Graphene and h-BN are exceptions, with $R_{MA/IL}$ values close to 1. Given the $R_{MA/IL}$ values are significantly higher than 1 for most 2D crystals, with similar values to that of Au films, we expect substantial adhesion interactions between those 2D layers and the Ag surface.

The $MoS_2$/Ag(111) interface is predicted to have an $R_{MA/IL}$ value of 1.70, which suggests that the $MoS_2$ crystal could be exfoliated using the Ag-assisted method. However, to further understand the binding characteristics and to ensure the exfoliation material does not impact the intrinsic properties of the 2D material, one also needs to consider the strength of the perturbation when Ag is interacting with the electronic structures of the 2D layers. To this extent, we further examine the related characteristics of this representative interface. Fig. 1b shows its differential charge density (DCD), from which significant covalent characteristics can be observed at the S/Ag interface, *i.e.* charge reduction near the interfacial atoms and charge accumulation between them. The binding energy of S/Ag (48 meV Å$^{-2}$; 0.39 eV per unit cell) is slightly stronger than that of S/Au (40 meV Å$^{-2}$; 0.35 eV per unit cell)[23]. Fig. 1c shows the unfolded band structures of the MoS$_2$/Ag interface, in which the shape and position of the valence (VB) and conduction bands (CB) of MoS$_2$ are almost unchanged in comparison to those of the freestanding MoS$_2$ monolayer. Therefore, the Ag thin film is even more conducive than the previously used Au thin film to maintain intrinsic electronic properties of those 2D materials laid on it.[23] Fig. 1d and 1e show the square of wave function norms of the fully occupied interfacial S-Ag bonding and anti-bonding states. Their energy levels are split by a few meV and depicts substantial interlayer attraction induced wave function overlap, known as covalent-like quasi-bonding, at the MoS$_2$/Ag interface. Given the sufficient 2D layer-Ag adhesive energy and the nearly unaffected electronic structures, our theory predicts that the Ag surface is, in principle, capable of offering an effective exfoliation strategy.

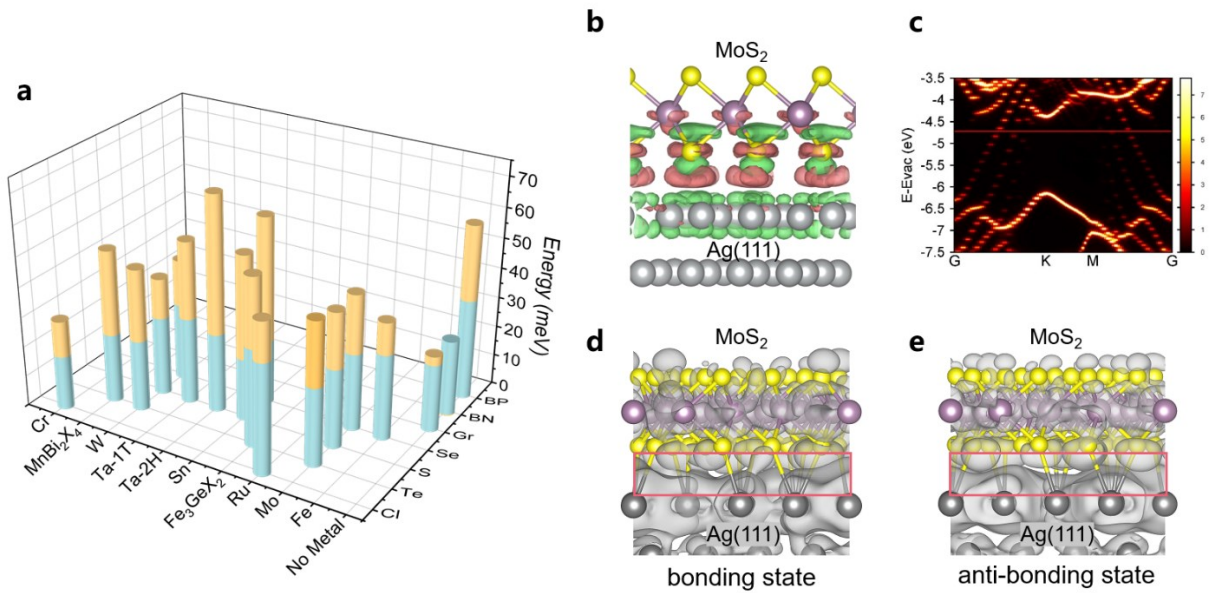

**Fig. 1: DFT calculations of 2D crystals on Ag substrate. a** Histogram of layered materials which shows the contrast of the interlayer binding energies (blue cylinders) and their adsorption energies on Ag (111) (yellow cylinders). The visible yellow cylinders represent the difference between the Ag/2D crystal and the interlayer interaction. **b** DCD plots of Ag (111)/MoS$_2$ interface. Isosurface value is 1× 10$^{-3}$ e Bohr$^{-3}$. **c** Unfolded band structure of Ag (111)/MoS$_2$ interface. The red line represents the Fermi level. **d-e** Isosurface contours of wavefunction norms for the bonding state and anti-bonding state using an isosurface of 6×10$^{-5}$ e Bohr$^{-3}$.

Surface roughness and plasmonic activation of Ag surfaces

To test whether the theoretical prediction is correct, we prepared Ag films with different thicknesses by thermal evaporation (details can be seen in Method part). Unlike evaporated Au that tends to form flat surfaces[23], evaporated Ag on a Si(111)[26], sapphire[27] or SiO$_2$[27, 28] substrate was demonstrated to have a rough surface, exhibiting nanoparticle-like structures (Fig. S1). These connected particle-like Ag can form nano cavities, which tend to promote the formation of localized-surface-plasmon-resonance (LSPR)[26, 27, 29],[27] or propagating SPP[26] (schematic shown in Fig. 2a). These two effects might substantially enhance PL intensities of several 2D semiconductor materials. A series of characterizations and simulations were thus conducted on the evaporated Ag thin films to unveil their plasmonic properties.

Fig. 2b depicts atomic force microscope (AFM) images of the surface morphology of an as evaporated 5 nm Ag film (left) and another Ag film (right) prepared using the template-stripping (TS) method (see Methods for details), respectively. The as evaporated 5 nm Ag film show a root-mean-square (RMS) roughness of 1.2 nm and the largest height variation is over 6 nm, while the surface prepared using the TS method yields a roughness of 0.4 nm, close to that of epitaxial-grown Ag films on Si(111) (0.3 nm)[26]. Statistical analysis revealed that such a rough surface can be divided into surface particles with lateral sizes ranging from 40 to 100 nm (mean ~80 nm), as a result of amorphous particle formation[23]. We can attribute this "particle agglomeration" phenomenon to that Ag does not wet with Ti, which leads to Vollmer-Weber-type growth and grain-boundary pining during Ag evaporation[28, 30, 31], which we directly illustrate in Fig. S2. As the deposition thickness increases from 5 nm to 25 nm, the surface morphology of those substrates shows similar characteristics in terms of particle shape, height and lateral size (Fig. S3).

The surface of rough Ag films has been reported to show broad plasmon resonances.[27] And when the exciton resonance matches with their plasmon resonances, PL intensity of supported samples will be enhanced. Relative reflectance spectra (ΔR/R) of the surfaces of those thin films were acquired to probe the plasmon resonance, where the reflectance is normalized to a reflective silver mirror. As shown in Fig. 2c, the 5 nm Ag films showed a broad plasmon resonance at ~2.1 eV, which can potentially be excited by a 532 nm laser source, and excitons generated in 2D crystals like monolayer $MoS_2$ and $WS_2$ (A-exciton resonances at ~1.86 eV and ~1.99 eV, respectively). For thicker substrates, their plasmon resonances generally experience a blue shift, which suggests that SPP excited in thicker Ag films can hardly couple with their exciton resonances. Furthermore, as calculated by finite-difference time-domain (FDTD) method (detailed modeling see Methods), Fig. 2d and 2e shows the top and cross-sectional view of field intensity distribution of Ag films centered at A-exciton resonance of monolayer $MoS_2$, respectively. For Ag films thinner than 9 nm, strong field confinement as well as long propagation of SPP (200 nm for 5 nm Ag films) can be achieved, indicating possible PL enhancement via Purcell factor and exciton-reexcitation, respectively. As Ag thickness increases, the propagation of SPP in in-plane directions decreases significantly (Fig. 2e and Fig. S4) and become negligible for films thicker than 9 nm. Our simulation shows that PL intensity of some 2D semiconductors, such as monolayer $MoS_2$ and $WS_2$, on thinner Ag films could be strongly enhanced. The enhancement, however, does start to decline as Ag thickness increases. The enhanced PL intensity suggested by simulation is confirmed with experimental results, and will be discussed in detail in the following sections.

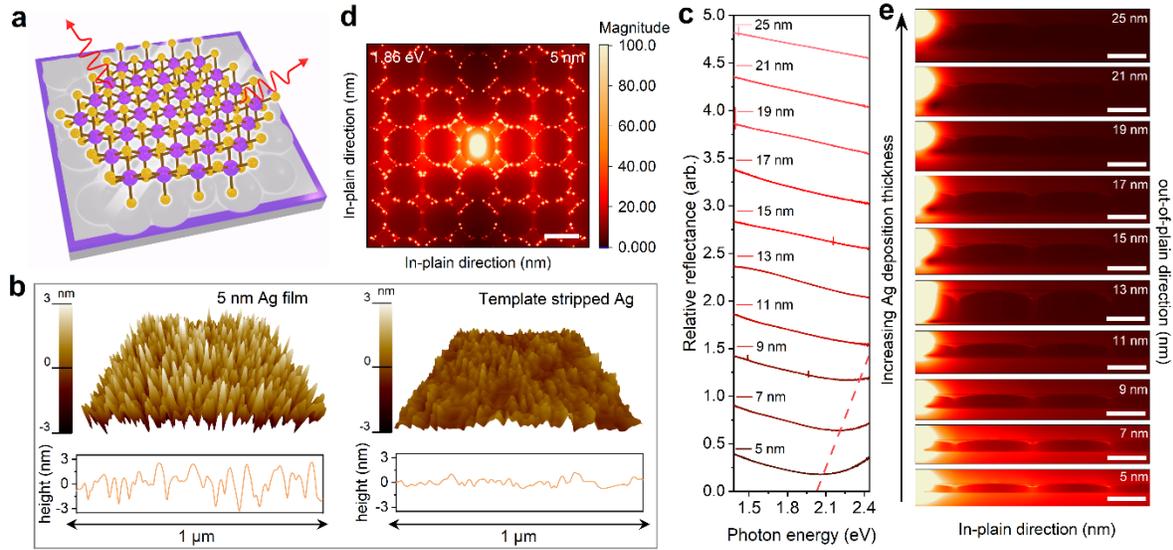

**Fig. 2: Surface and optical characterizations of Ag films. a** Schematic illustration of a monolayer MoS$_2$ exfoliated onto the surface of a 5 nm Ag film. **b** AFM of 5 nm Ag film surface (left) and TS Ag/epoxy/glass slide (right). **c** Relative reflectance spectra ΔR/R of Ag films with varying thicknesses acquired in contrast to a self-made silver mirror. Red dashed line denoting the position for plasmon resonance is employed for clarity. **d** FDTD simulated electric field intensity distribution (top view) in the case shown in **a** centered at 1.86 eV, indicating that the Ag films facilitate strong field confinement as well as long propagation of SPP. Scale bar is 80 nm. **e** FDTD simulated electric field intensity distribution (cross section) in the case shown in **a** centered at 1.86 eV with thickness from 5 nm to 25 nm, the distance of SPP propagation along the X axis exhibit a conspicuous decline as thickness of Ag films increase. Scale bar is 40 nm.

## Exfoliation of plasmonic activated 2D monolayer crystals

Based on theoretical and simulation results, we experimentally explored using Ag films as exfoliation medium for layered materials. Here, 12 representative 2D monolayers were successfully exfoliated using this Ag-assisted exfoliation method. Fig. 3a shows the schematic illustrations of Ag-assisted exfoliation, which was also tested and confirmed to be compatible on various substrates, such as SiO₂/Si (Fig. 3b), sapphire (Fig. 3c), PET (Fig. 3d) and quartz surfaces. Those monolayers can also be exfoliated onto Ag surfaces using the TS method (Fig. 3e, detailed preparation is discussed in Methods), which indicates that surface roughness is not a critical issue for successful exfoliation. Notably, the yield of monolayer drastically drops after exposure of the as-prepared Ag surface to air in 10 seconds (Fig. S5), indicating the ambient atmosphere condition is detrimental to Ag-assisted exfoliation (most likely oxidation).

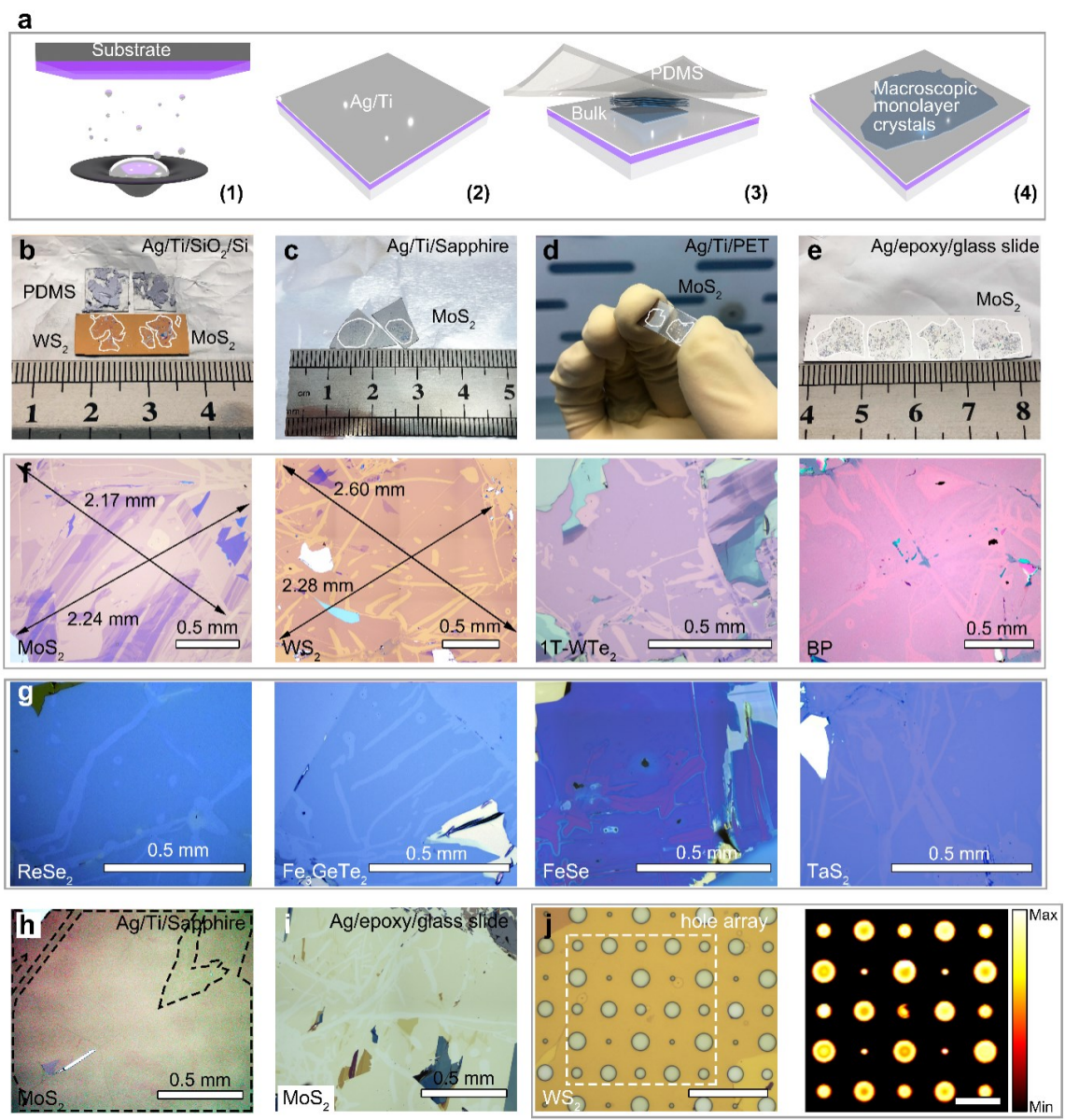

**Fig. 3: Ag-assisted exfoliation procedures and optical characterization of exfoliated samples. a** Schematic illustration of the exfoliation procedures. **b** Exfoliated macroscopic $MoS_2$ and $WS_2$ on 15 nm Ag film supported by $SiO_2$/Si substrates, and bulk crystals on PDMS tapes. **c** Exfoliated macroscopic $MoS_2$ supported by sapphire substrates. **d** Exfoliated $MoS_2$ supported by elastic PET substrates. **e** Exfoliated $MoS_2$ on Ag/epoxy/glass slide substrate. **f** Optical microscope images of some 2D crystals exfoliated on 15 nm Ag film, including $MoS_2$, $WS_2$, 1T-$WTe_2$, and BP. **g** Optical microscope images of exfoliated millimeter size 2D crystals on 5 nm Ag film, including $ReSe_2$, $Fe_3GeTe_2$, FeSe, and $TaS_2$. **h-i** Optical microscope images of exfoliated millimeter size $MoS_2$ on sapphire substrate and TS Ag, respectively. **j** Optical microscope and PL mapping images of exfoliated monolayer $WS_2$ on 15 nm Ag film with hole array, the scale bars in the two images are 40 μm and 20 μm, respectively.

Fig. 3f, 3g, S6 and S7 illustrate those 12 exfoliated monolayers, including $MoS_2$, $MoSe_2$, $MoTe_2$, $WS_2$, $WSe_2$, $ReS_2$ and BP (semiconductors), 1T-$WTe_2$ and FeSe (superconductors), $Fe_3GeTe_2$ and $MnBi_2Te_4$ (magnets) and $TaS_2$ (CDW), which show macroscopic size and high uniformity. Furthermore, the presented method, behaves similar to Au-enhanced exfoliation in that it does not require the substrate to be fully covered by Ag. This enables direct exfoliation[32] of free-standing 2D monolayers using patterned hole-array substrates. Fig. 3j shows an optical microscope image and a corresponding PL mapping of a large-scale suspended $WS_2$ monolayer. The suspended $WS_2$ monolayer presents much stronger PL emission on hole areas than those supported areas. Additionally, low wavenumber Raman modes of 2D crystals suspended and supported were acquired, the suppression of breathing and shearing Raman modes for samples on Ag films implies the existence of CLQB as demonstrated in Fig. S8 and S9. This CLQB induced pinning effect is also found in Au/2D materials interfaces[32, 33].

## Observation and mechanism of PL enhancement

For most 2D semiconductors, PL intensities of SiO$_2$ supported monolayers are typically weaker than those of suspended monolayers[34], which are further suppressed in metal supported monolayers due to additional charge transfer[35] from the metal surfaces. As discussed above, the yield of exfoliated 2D flakes is not sensitive to the roughness of Ag. However, the roughness strongly affects plasmonic properties of exfoliated 2D monolayers. As expected, no appreciable PL intensity was observed on the TS Ag film supported MoS$_2$ monolayer because of the flat metallic surface (black, Fig. 4a). To our surprise, an extraordinarily prominent A-exciton emission located at ~1.86 eV (PL intensity and peak position mapping shown in Fig. S10) was observed on the rough surface Ag supported monolayer (orange, Fig. 4a). The PL intensity of monolayer MoS$_2$ on the rough Ag film is 3.1, 11 and over 200 folds stronger than those of suspended, SiO$_2$ and TS Ag supported monolayers, respectively. Fig. 4b summarizes the comparison of PL spectra for suspended and rough Ag surface supported MoS$_2$, MoSe$_2$, WS$_2$ and WSe$_2$ monolayers. Similarly, the PL emission of plasmonic coupled MoSe$_2$ is also 4.8 times than that of the suspended MoSe$_2$. However, unlike those molybdenum-based samples, the supported WS$_2$ and WSe$_2$ monolayers on rough Ag film does suffer from significant PL quenching. Their PL intensities are 1/24 and 1/620 times those of WS$_2$ and WSe$_2$ monolayers exfoliated onto the SiO$_2$/Si substrate, respectively. All PL measurements were conducted under the exact circumstances, details see Methods.

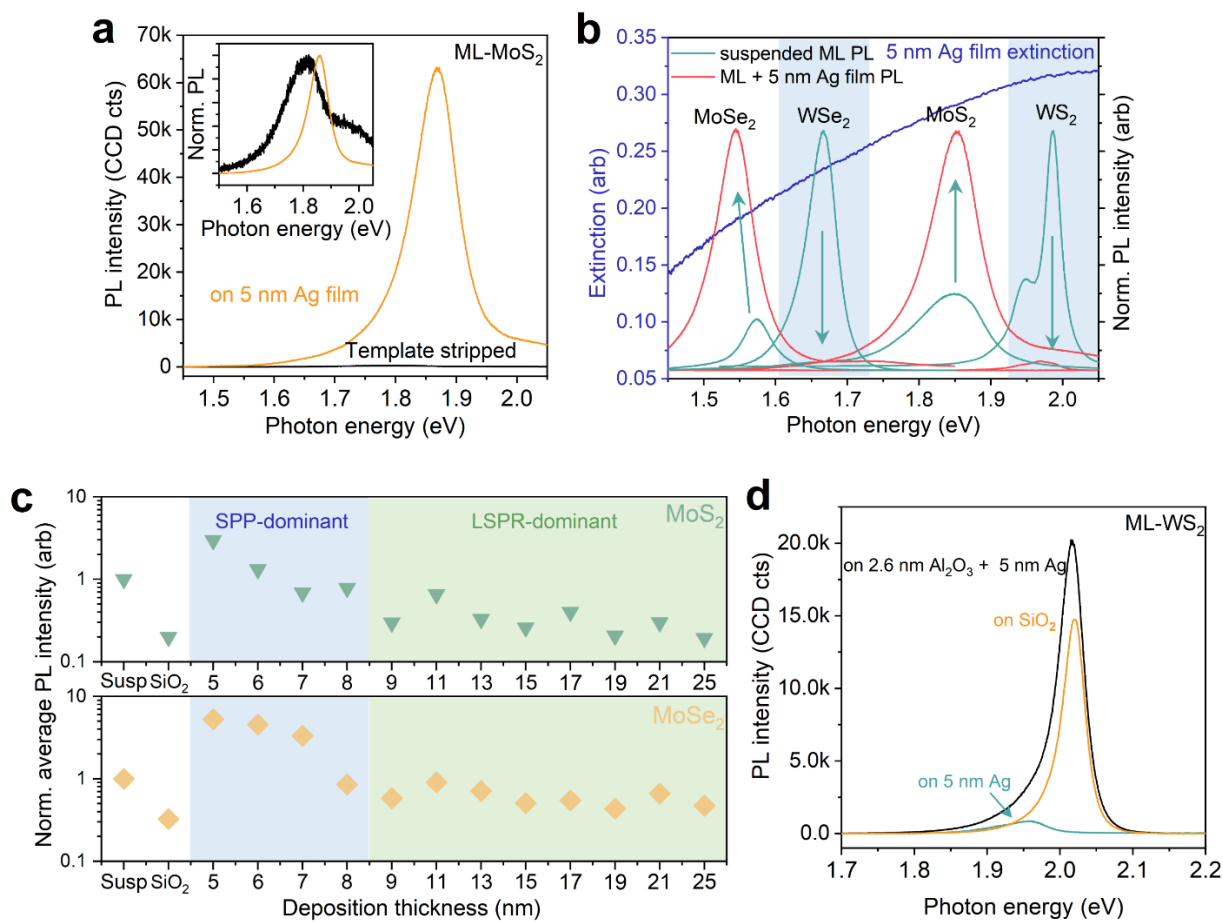

**Fig. 4: PL behavior of four exfoliated TMDC semiconductors on Ag film. a** PL spectra of monolayer MoS$_2$ exfoliated onto 5 nm Ag films and TS Ag substrates. Inset: normalized PL spectra of monolayer MoS$_2$ exfoliated onto 5 nm Ag films and TS Ag. **b** Extinction spectrum of 5 nm Ag film (blue) and normalized PL intensity of monolayer MoS$_2$, MoSe$_2$, WS$_2$, and WSe$_2$ on 5 nm Ag films (orange) and on hole arrays (green). **c** Normalized average PL intensity of as-exfoliated monolayer MoS$_2$ and MoSe$_2$ on Ag films with different deposition thickness and SiO$_2$/Si substrates, normalized to PL intensity of their respective suspended monolayers, each average intensity data was acquired at 5 different spots. Generally, PL emission of Ag supported samples experience a decrease with increasing Ag thickness. **d** PL spectra of monolayer WS$_2$ on 5 nm Ag film (green), on SiO$_2$/Si substrate (orange), and 2.6 nm Al$_2$O$_3$ + 5 nm Ag film (black), which indicates PL quenching is mainly contributed by charge transfer.

The three orders of magnitude variation of PL intensity is, most likely, a result of the competition between exciton-plasmon coupling activated PL enhancement[31, 36] and Ag induced quenching effect [27, 37, 38]. The exciton-plasmon coupling usually leads to propagation of SPP[26] so that the exciton-SPP-photon conversion and exciton re-excitation could enhance PL intensity for 2D semiconductors. An additional effect lies in the enhancements of incident excitation (near-field emission) and excitonic emission (spontaneous emission), which are positively-related to the local photonic densities[27, 29, 39] and highly dependent on the surface morphology of Ag films (*e.g.* roughness)[27, 29, 40].

We examined the thickness dependent competition of these two competing effects by exfoliating $MoS_2$ and $MoSe_2$ samples on a series of Ag films in a range of thickness from 5 nm to 25 nm. Fig. 4c presents thickness-dependent normalized PL intensities of monolayer $MoS_2$ ($MoSe_2$) on the series of Ag films (those for $WS_2$ and $WSe_2$ were shown in Fig. S11), in which the strongest PL was found on the 5 nm sample. The intensity clearly decays from 5 to 8 nm and remains weak for thicker ones. The critical thickness of 8 nm is comparable with the simulated critical thickness of 9 nm for propagating SPP[26], supporting its dominant role in strengthening the PL intensity of $MoS_2$ on Ag films thinner than 8 nm. For Ag layers thicker than 9nm, although the propagation of SPP is suppressed, the thickness independent LSPR still contributes towards the PL intensity and provides a small intensity enhancement.[27]

For $WS_2$ monolayer, its Fermi level sits at −4.87 eV[41], over 0.6 eV lower than that of the Ag(111) surface (−4.3 eV), giving rise to a n-type doping to $WS_2$ at the $WS_2$/Ag interface. The n-doping promotes the formation of negatively-charged trions and thus lowers its PL quantum yield,[25, 26, 28, 42, 43] which is even more efficient in high quantum yield materials (*e.g.* monolayer $WS_2$ and $WSe_2$).[27, 38] To verify this effect, a 2.6 nm $Al_2O_3$ dielectric layer was deposited by atomic layer deposition (ALD) onto the 5 nm Ag film, where a flake of monolayer $WS_2$ was then transferred. As expected, the PL emission of monolayer $WS_2$ on 5 nm Ag with the $Al_2O_3$ dielectric layer is 1.3 times stronger than the sample on the $SiO_2$/Si substrate (Fig. 4d), which confirms that PL quenching primarily originates from electron transfer of Ag to $WS_2$. Such quenching effect is also expected for other high work function monolayers, *e.g.* $WSe_2$.

## Conclusion

We have presented a universal Ag-assisted exfoliation technique that creates a unique integration between large-scale 2D crystals and plasmonic nanostructures, while offering multiple advantages over Au-enhanced exfoliation. The effectiveness and universality of this technique is justified by theoretical calculations as well as spectroscopy results, which confirmed the strong interactions caused by CLQB exist between Ag and 2D materials, exceeding the interlayer van der Waals interactions. Optical characterizations were conducted immediately after exfoliation, showing enhanced PL emissions of monolayer $MoS_2$ and $MoSe_2$, as a result of SPP propagation enhanced light-matter interaction. The high quality and yield of as-exfoliated samples proves that our technique is not only important for basic research of emergent 2D materials, but also shows great potential in industrial production of ultrathin 2D crystals. This study provides an alternative option for direct exfoliation and optical characterization of emergent 2D crystals. Whether large-scale 2D crystals can be directly exfoliated onto other metal substrates for functionalities such as SERS or fano resonances, or even insulating substrates where their intrinsic physical properties are well kept and readily to be explored are worth for further investigations.

## Methods

DFT calculations: (DFT calculations)

DFT calculations were performed using the generalized gradient approximation for the exchange-correlation potential, the projector augmented wave method,[44, 45] and a plane-wave basis set as implemented in the Vienna ab initio simulation package (VASP).[46] Van der Waals interactions was performed at vdW-DF level for all calculations, with the optB86b functional for the exchange potential.[47] The energy cutoff for the plane-wave basis set was set to 700 eV for variable volume structural relaxation of pure materials and 500 eV for invariant volume structural relaxation of 2D materials on Ag (111) surface. A k-mesh of $9 \times 9 \times 1$ was adopted to sample the first Brillouin zone of the conventional unit cell of the Ag (111) slab. The mesh density of k points was kept fixed when calculating other structures. Seven layers of Ag atoms, separated by an 18 Å vacuum region, were employed to model the Ag (111) surface. In geometry optimization, all atoms in the supercell except for the last four layers of Ag atoms were allowed to relax until the residual force per atom was less than 0.01 eV Å$^{-1}$. To make the electronic properties more accurate, lattice parameters of 2D materials were fixed and those of the Ag (111) surface slab were changed, with the lattice mismatch between 2D layer and the Ag (111) surface being kept lower than 4.5%. Differential charge density (DCD) was calculated using $\Delta\rho_{DCD} = \rho_{All} - \rho_{Ag} - \rho_{2D}$. Here $\rho_{All}$ is the total charge density of the 2D layer/Ag (111) interface, while $\rho_{Ag}$ and $\rho_{2D}$ are the total charge densities of the individual Ag surface and the 2D layer, respectively. Unfolded band structure was calculated using the KPROJ program based on the k-projection method.[48, 49]

## FDTD simulations

The FDTD simulations were performed on a commercial software. The defined dielectric constant of Ag was from Johnson and Christy. Semi-spheroid Ag nanoparticles (5 nm height and 80 nm in lateral size) placed on Ag films with different thickness were used to mimic the surface morphologies of as-deposited Ag films and to qualitatively analyze the propagation of SPP. A 0.6 nm thick monolayer MoS$_2$ asymmetric thin film (in plane and out of plane dielectric constants are 2.22 and 1.70) is placed on top of the Ag nanoparticles, and an in-plane dipole is placed at the center of the film to mimic the effect of an exciton.

## Ag-assisted mechanical exfoliation

The SiO$_2$/Si substrate was first treated with pure oxygen plasma or air plasma for approximately 5 mins, and at a power of 40 W, for cleansing the substrate surface of contaminations and ensuring better adhesion between substrate and metal layer. The metal layer deposition was carried out in an in-glovebox thermal evaporation system (VNano). An ultrathin adhesive layer was first evaporated (0.3 Å s$^{-1}$) on SiO$_2$/Si substrate (100 nm oxide), followed by deposition of Ag film (1 Å s$^{-1}$). After metal deposition, an as-cleaved surface of bulk crystals on PDMS (Gel-Pak) tape was brought into contact with the substrate. After gently pressing the PDMS tape vertically for approximately 1 min, uniform contact can be ensured and tape can then be instantly removed from the substrate. Millimeter size or even bare eyes observable monolayers can thus be obtained with ultraclean surfaces, of which size is merely restrained by the size of our bulk crystals. Nonetheless, exfoliation yield still can be severely hindered if the adhesive is too thin. Once the Ti layer is less than 1 nm and Ag layer is less than 5 nm, the yield of monolayer drops steeply. This is probably because 1 nm Ti is not a continuous film that can no longer offer sufficent adhesion with the supporting below. In this case, evaporated Ag/Ti can be easily peeled off from the substrates. With optimized 15 nm Ag/3 nm Ti films, success rate of exfoliation is nearly unity.

The bulk crystals are mainly bought from commercial companies (2D Semiconductors and HQ Graphene), some of them are supplied by other groups. All of these crystals can be exfoliated into large-scale monolayers with size dependent on the size of our bulk crystals. The thickness of bulk crystals is one critical factor to our exfoliation. Once the bulk crystal is too thick (thicker than 0.5 mm), its rigidness will influence a fine contact between itself and the substrate, which decreases or even eliminate the size of yielded monolayers. There is no need to cleave the bulk crystals too thin because they might be reduced into small fragments. The optimized bulk crystals should be continuous and thin, yet not transparent if observed against light illumination. This exfoliation is not fitted for ambient conditions that the exposure to humidity and oxygen can cause fast oxidation of surficial Ag, and consequently loses all adhesion ability. However, it can be well performed in a glovebox with humidity and oxygen concentration < 0.01 ppm, where the as-prepared adhesive layer can last for several days and still keep its effectiveness. We also tested that under 100 ppm oxygen concentration, as-deposited Ag films lose its functionality in roughly 6 hours.

For preparation of suspended samples, $SiO_2$/Si wafer was first patterned by UV lithography (Karl Süss MA6), and then hole arrays were opened up by reactive ion etching ($CF_4/O_2$) with 10 μm diameter and 10 μm depth. After that, ultrathin Ag and Ti were deposited in an in-glovebox thermal evaporation system, where the same Ag-assisted exfoliation technique illustrated above was conducted.

In order to find other exfoliation media layer, we also tested copper (Cu), chromium (Cr) and titanium (Ti) for exfoliation of several 2D crystals. The tested layered crystals include $MoS_2$, $WS_2$ and BP. However, the flake sizes are very small and the yield is quite low, similar as the 2D flakes exfoliated on bare $SiO_2$ substrates.

## Template stripping of Ag film

150 nm Ag was first evaporated in an in-glovebox thermal evaporation system onto a sacrificial silicon substrate at 1 Å s$^{-1}$ rate to reduce the possibility of grain boundary formation. The wafer was then cut into specified sized pieces and coated with moderate epoxy adhesive. Glass slides were then pressed onto the epoxy so that the epoxy can cover their whole substrate surfaces. Afterwards, glass slide/epoxy/Ag/$SiO_2$/Si stacks were heated up on a hotplate at 60 °C for 2 hours. Then glass slide/epoxy/Ag can be easily peeled off from the $SiO_2$/Si sacrificial layer, which yielded a considerably smooth Ag surface.

## Preparation of suspended samples

SiO$_2$/Si wafer was first patterned by UV lithography (Karl Süss MA6), and then hole arrays were opened up by reactive ion etching (CF$_4$/O$_2$) with 10 μm diameter and 10 μm depth. After that, ultrathin Ag and Ti were deposited in an in-glovebox thermal evaporation system, where the same Ag-assisted exfoliation technique illustrated above was conducted.

## Optical and surface characterization

The Raman and PL measurements were operated on a WITec alpha300R system equipped with a wavelength 532 nm diode-pumped solid-state laser and power at 0.8 mW (if not otherwise noted). The low wavenumber Raman detection capability of this device is > 7 cm$^{-1}$. The silicon Raman mode at 520.7 cm$^{-1}$ was used for calibration before measurements.

Surface characterizations were carried out using AFM (Oxford, Asylum Research Cypher S) in a tapping mode.

## Contributions

Q.F., J.Q.D. and X.Y.H. contributed equally to this work. Y.L.W., Z.H.N., W.J. and Y.H. are equally responsible for supervising this discovery. Y.H., W.J., and Q.F. conceived the project. J.Q.D., Y.H.P., and W.J. performed DFT calculations. Q.F. prepared exfoliated 2D crystals, and performed optical measurements. Q.F., J.P.L and L.L.Y. acquired extinction spectra of as-deposited Ag films, and performed FDTD simulations. Q.F., X.Y.H, X.H. and Z.Y.S. measured the surface morphology of as-deposited Ag films and TS Ag films by AFM. Q.F., Y.Y.D., Y.H., W.J., J.Q.D., L.L.Y., X.J.Z. and Y.H.P. analyzed the data, wrote the manuscript, and all authors discussed and commented on it.

# Supplementary Information for

# One-step exfoliation method for plasmonic activation of large-area 2D crystals

Qiang Fu†, Jia-Qi Dai†, Xin-Yu Huang†, Yun-Yun Dai, Yu-Hao Pan, Long-Long Yang, Zhen-Yu Sun, Tai-Min Miao, Meng-Fan Zhou, Lin Zhao, Wei-Jie Zhao, Xu Han, Jun-Peng Lu, Hong-Jun Gao, Xing-Jiang Zhou, Ye-Liang Wang*, Zhen-Hua Ni*, Wei Ji* and Yuan Huang*

†These authors contributed equally to this work.

Correspondence to: yeliang.wang@bit.edu.cn (Y.L.W.), zhni@seu.edu.cn (Z.H.N.), wji@ruc.edu.cn (W.J.), and yhuang@bit.edu.cn (Y.H.)

**Table of Contents**

1. Surface morphology characterizations for Ag and Au films. (Fig. S1-S3)
2. FDTD simulation of electromagnetic field distributions of monolayer $MoS_2$/Ag film hybrid structures. (Fig. S4)
3. Raman, PL and electrical characterizations for the exfoliated 2D materials, heterostructures and suspended 2D materials on Ag film. (Fig. S5-S12)
4. Calculated energies of 2D materials supported by Ag(111). (Table S1)

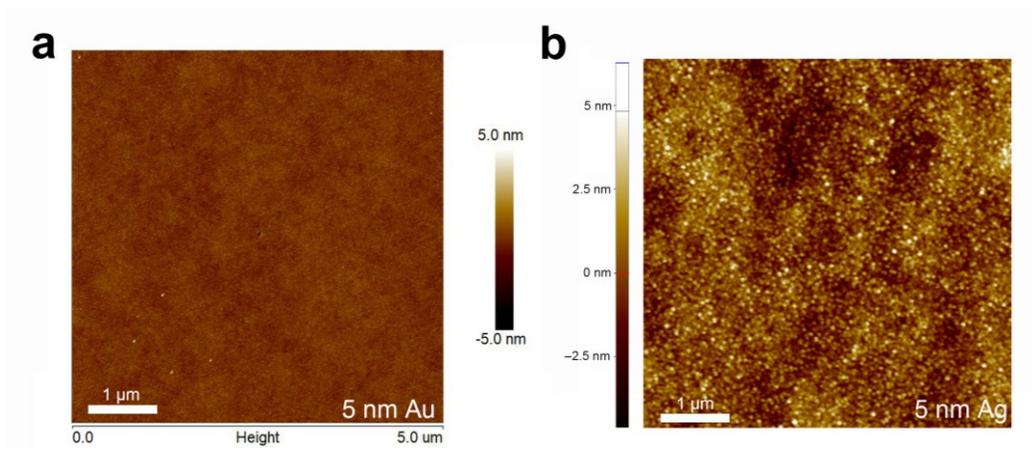

Fig. S1 **a-b** AFM images of 5×5 μm area 5 nm Au/2 nm Ti film (a) and 5 nm Ag/2 nm Ti film on 300 nm SiO$_2$/Si substrate.

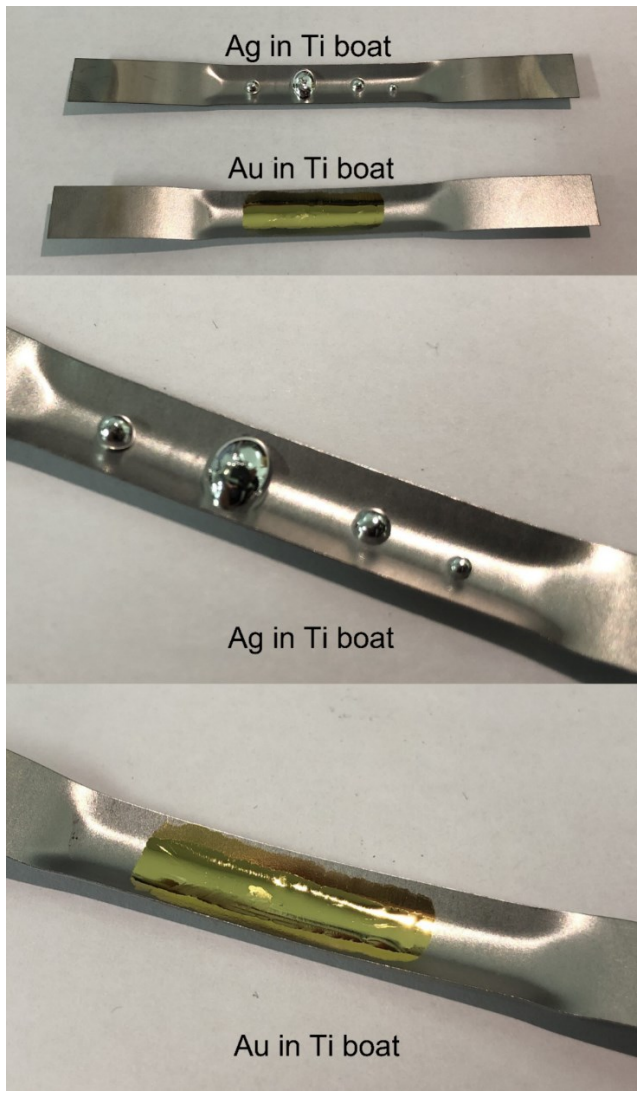

Fig. S2 Au and Ag in tungsten boat covered with 5 nm thick evaporated Ti after solidification from liquid states. The balled-up Ag implies obvious dewetting between Ag and Ti, whereas Au wets Ti and tends to form uniform films.

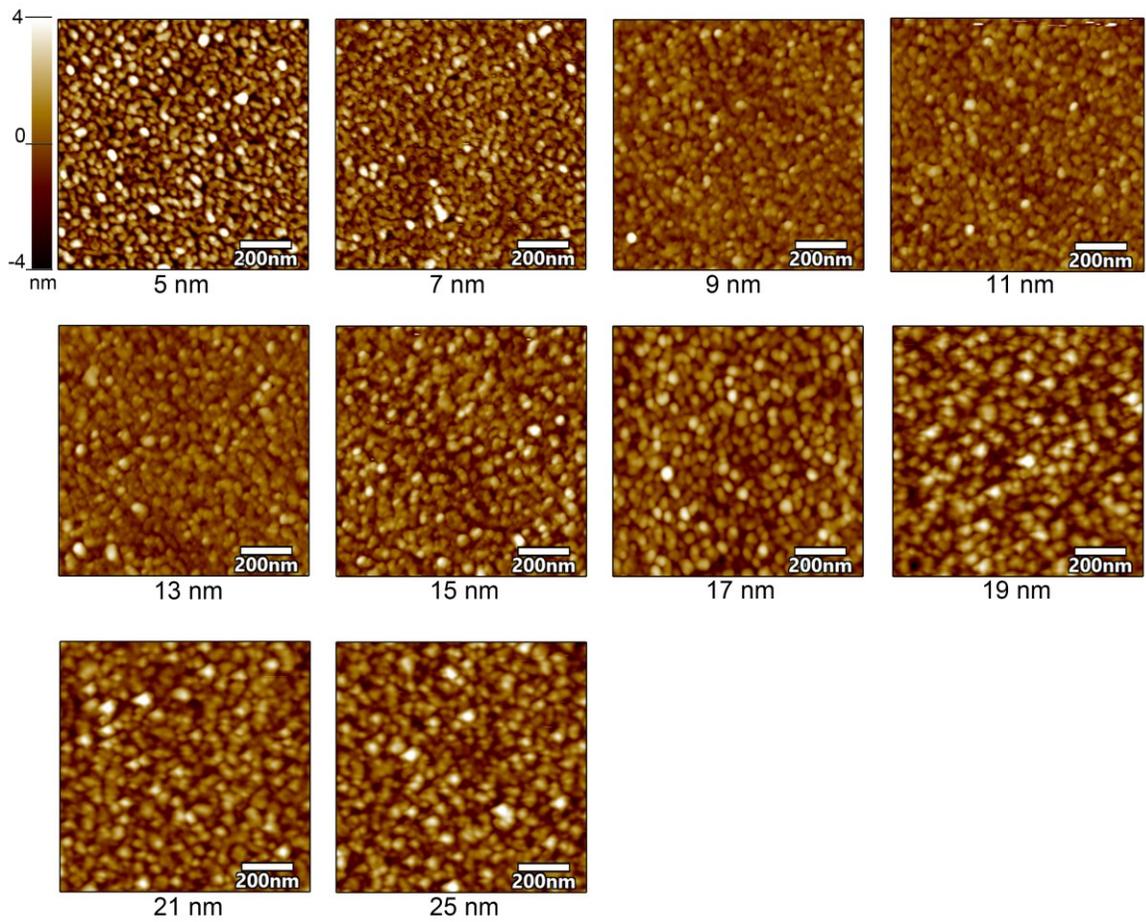

Fig. S3 AFM images showing the surface morphology of Ag films with increasing thickness, as the deposition thickness increases, the surface morphology barely changes.

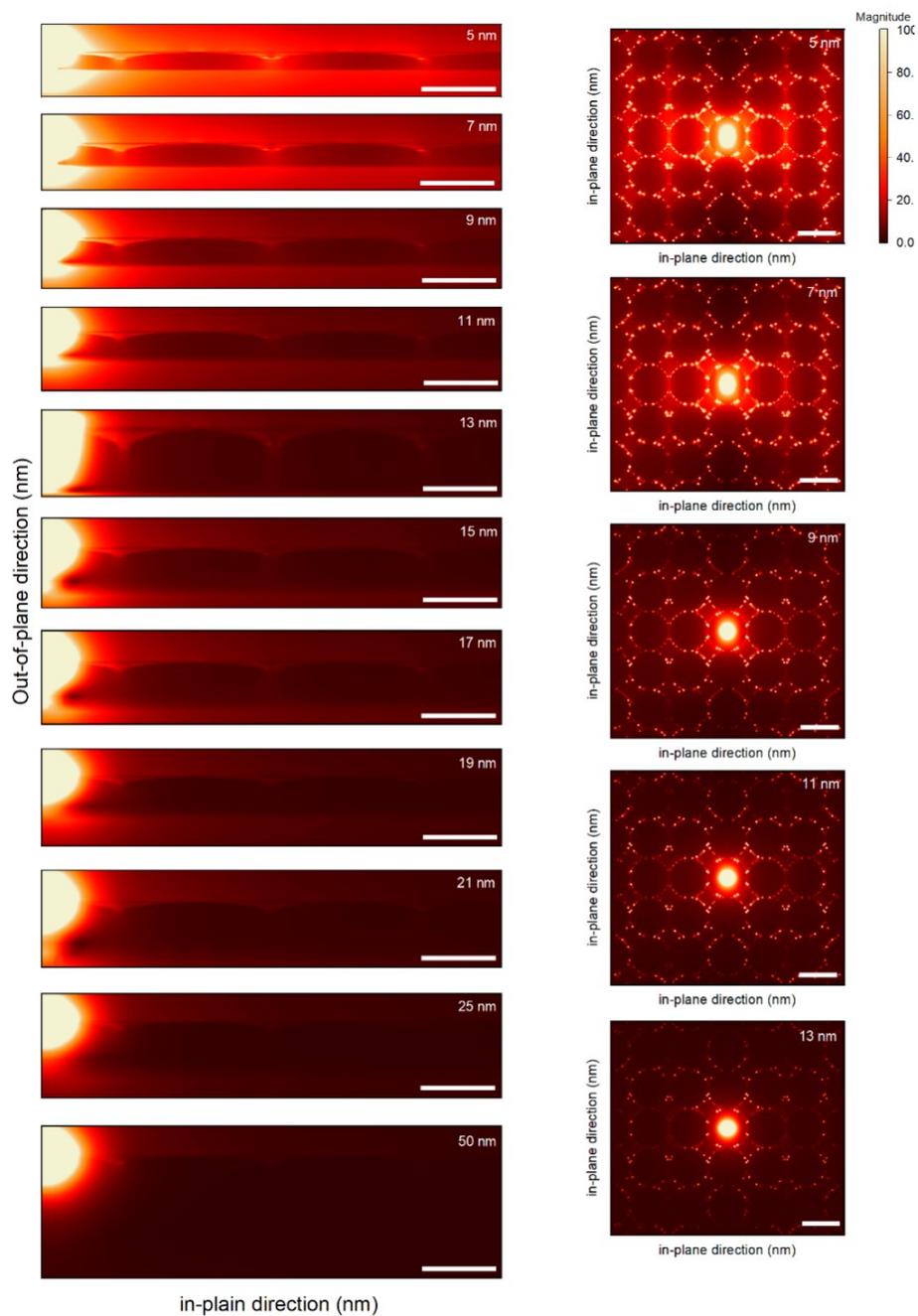

Fig. S4 FDTD simulation of electromagnetic field distribution in the in-plane and out-of-plane direction of monolayer $MoS_2$/Ag film hybrid structures with different deposition thickness for photon energy at 1.86 eV (A exciton emission of monolayer $MoS_2$), scale bars are 40 nm and 80 nm, respectively. The propagation of SPP severely decreases with increasing Ag deposition thickness, indicating less photons are coupled into the SPP modes.

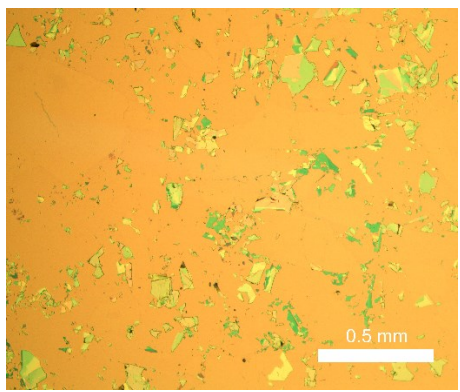

Fig. S5 Optical microscope image of MoS$_2$ exfoliated onto 15 nm thick Ag after exposing to air for approximately 10 seconds, no macroscopic monolayers can be found.

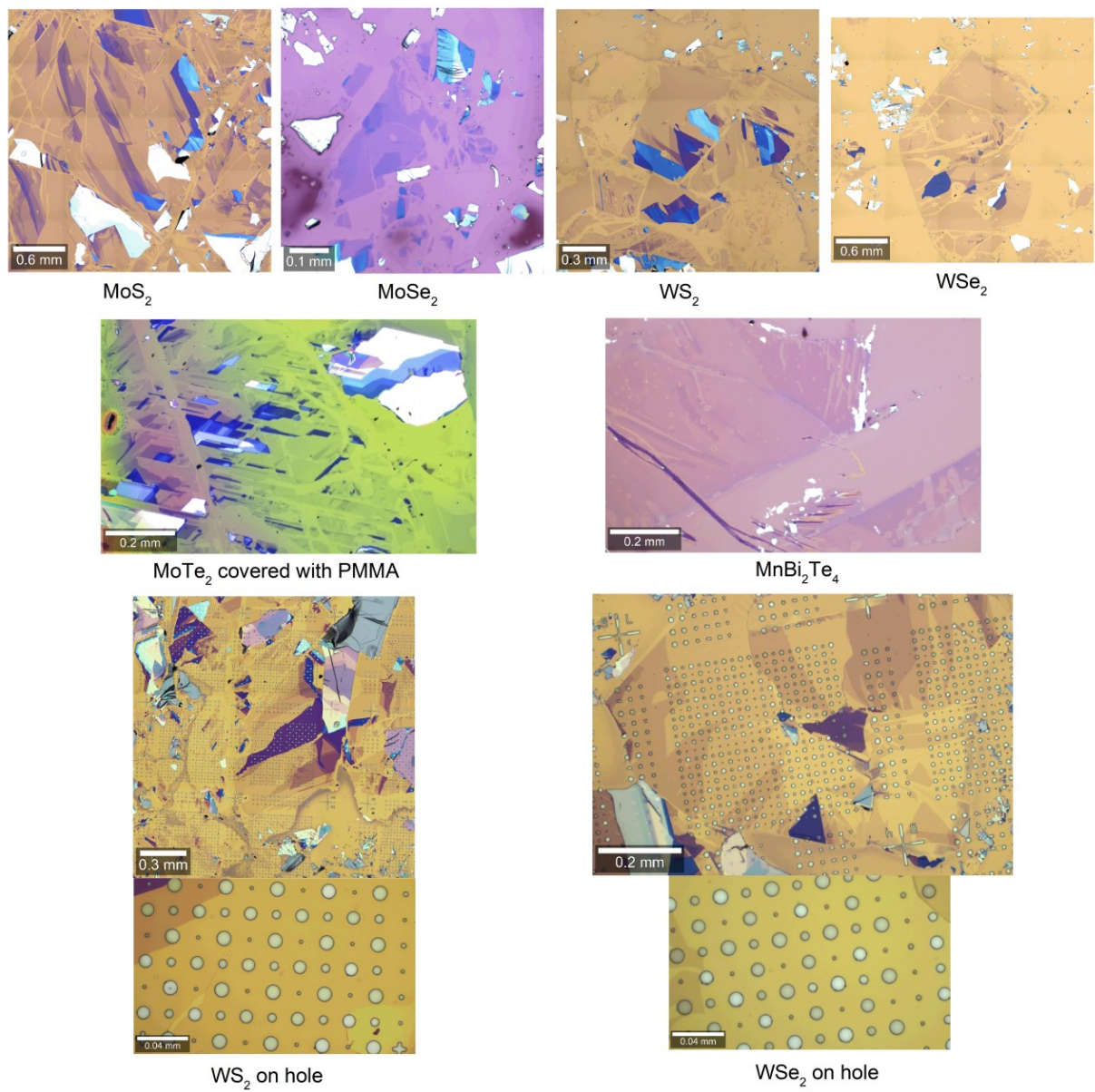

Fig. S6 Optical microscope images of large-scale monolayer or few-layer $MoS_2$, $MoSe_2$, $WS_2$, $WSe_2$, $MoTe_2$ and $MnBi_2Te_4$ exfoliated onto Ag and patterned Ag films.

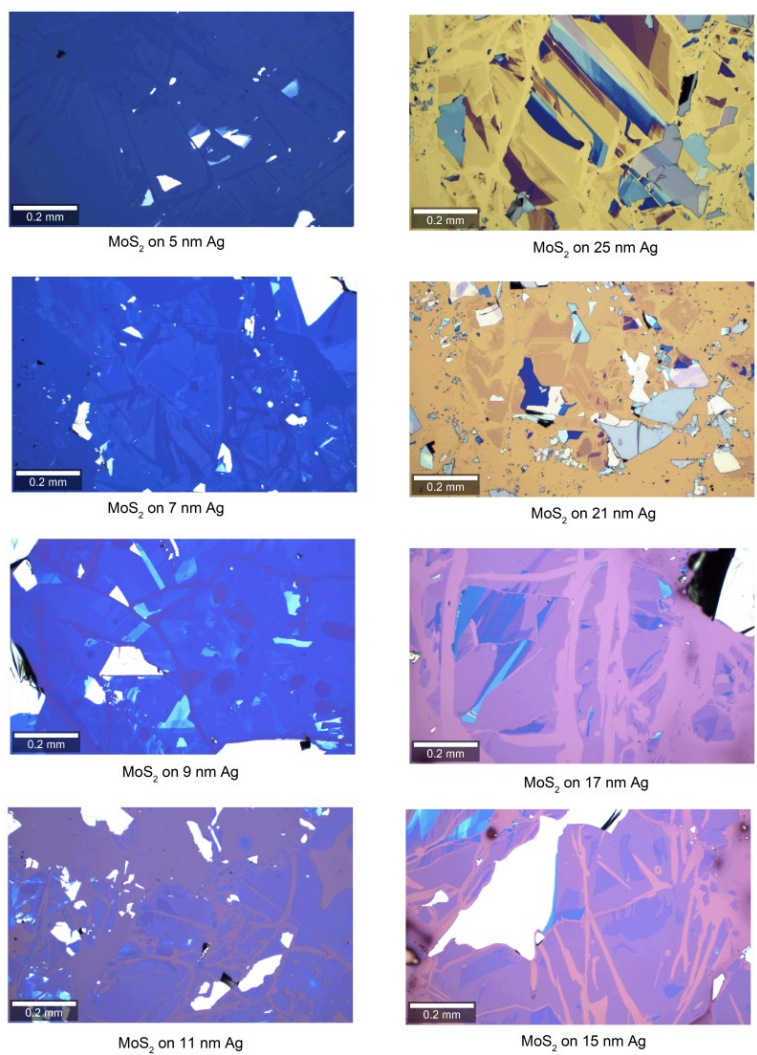

Fig. S7 Optical microscope images of large-scale monolayer or few-layer MoS$_2$ exfoliated onto Ag film with varying thickness.

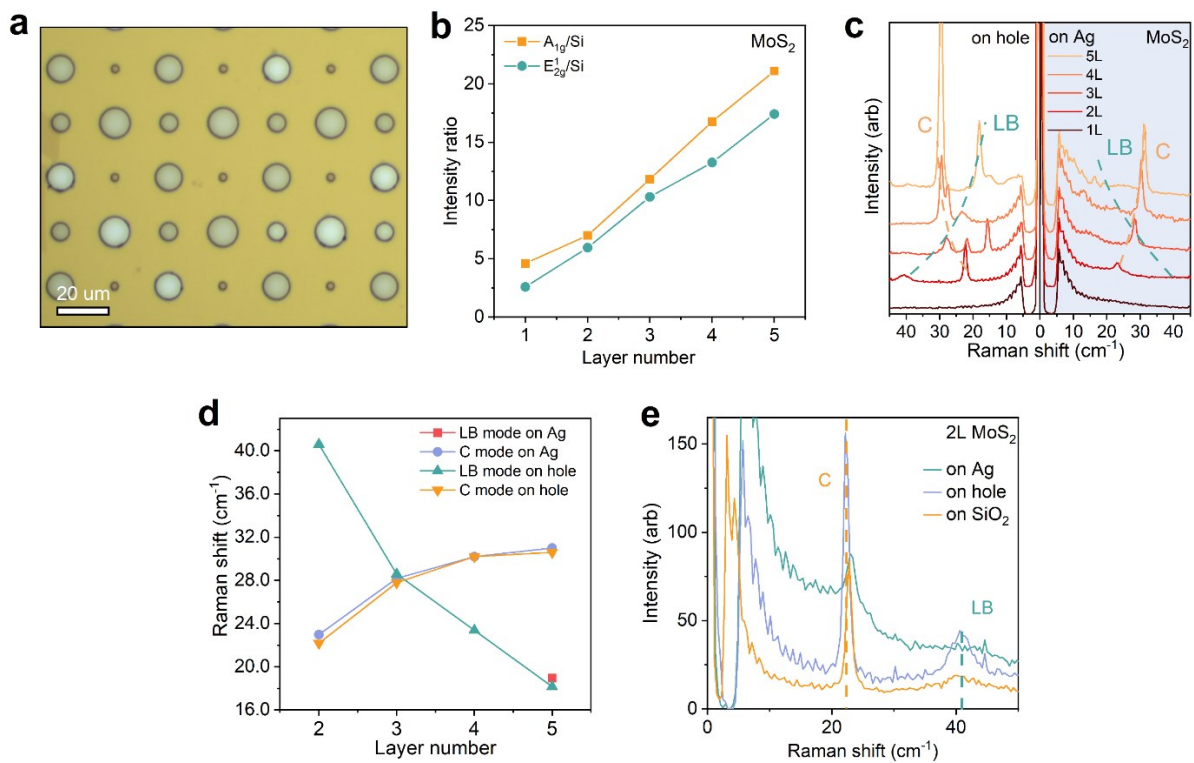

Fig. S8 Raman spectra of 2D crystals on Ag film. **a** Optical microscope image of as-exfoliated free-standing monolayer $MoS_2$ on 15 nm Ag film. **b** Evolution of intensity ratio between characteristic Raman modes of 1L to 5L $MoS_2$ on 15 nm Ag films and Si around 520.1 $cm^{-1}$. **c** Comparison between 1L to 5L $MoS_2$ exfoliated onto hole arrays (left) and onto 15 nm patterned Ag film (right), suppression of interlayer vibrational modes implies strong interactions between Ag layer and as-exfoliated samples. **d** Peak positions of LB and C modes derived from **c**, LB modes of 2L to 4L $MoS_2$ on Ag are not presented because they cannot be distinguished. **e** Comparison of LW Raman modes of 2L $MoS_2$ on Ag, hole and $SiO_2$/Si, dashed lines are marked for shearing mode (C) and layer breathing mode (LB), respectively.

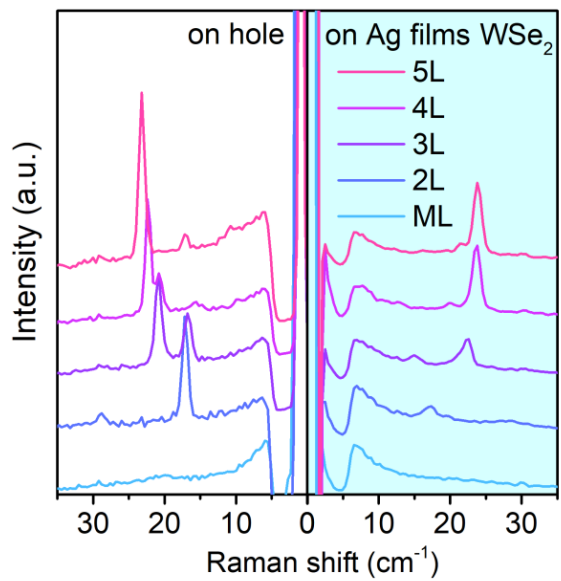

Fig. S9 Comparison of LW Raman activated modes between 1L to 5L WSe$_2$ exfoliated onto hole arrays (left) and onto 15 nm Ag films (right), suppression of interlayer vibrational modes implies strong interactions between Ag films and as-exfoliated samples.

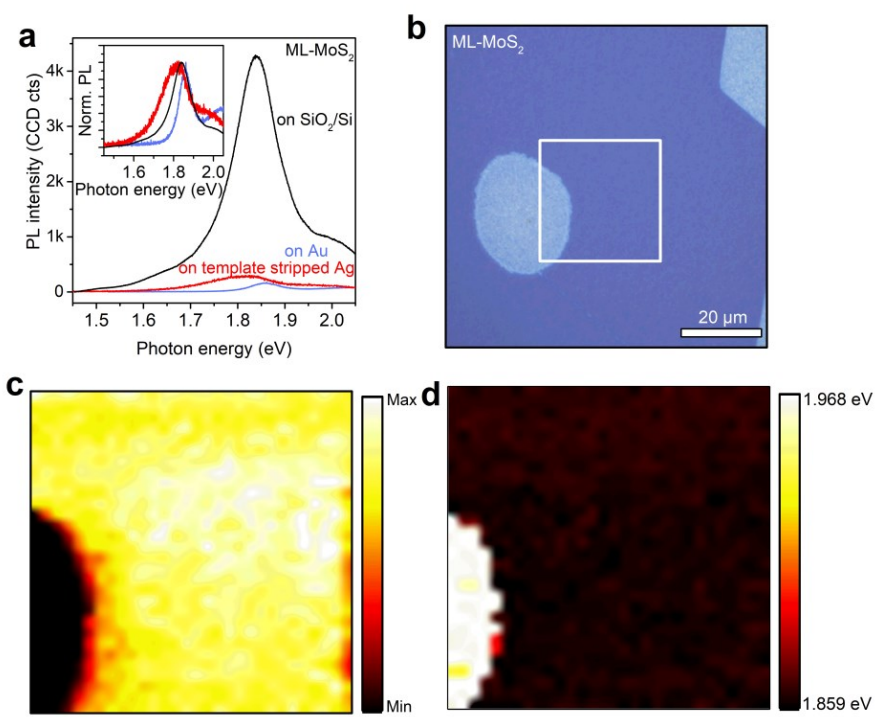

Fig. S10 **a** Comparison of PL intensity of monolayer $MoS_2$ exfoliated onto 5 nm Au (red), template stripped Ag (blue), and $SiO_2$/Si (black) substrates. **b** Optical microscope image of as-exfoliated monolayer $MoS_2$ on 5 nm APSs, the surface roughness is observable to the bare eye, rectangle area denotes the 30 x 30 μm mapping area. **c** PL intensity mapping of the as-exfoliated monolayer $MoS_2$, extraordinarily strong PL is nearly uniform across the whole area. **d** PL peak position mapping of as-exfoliated monolayer $MoS_2$, which is almost invariable at ~1.86 eV.

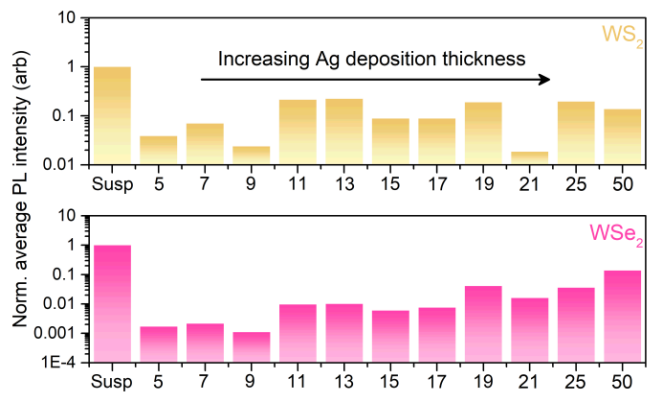

Fig. S11 Normalized average PL intensity of as-exfoliated monolayer $WS_2$ and monolayer $WSe_2$ on Ag films with increasing deposition thickness, normalized to PL intensity of their respective suspended monolayers, each average intensity data point was the average intensity at 5 different spots.

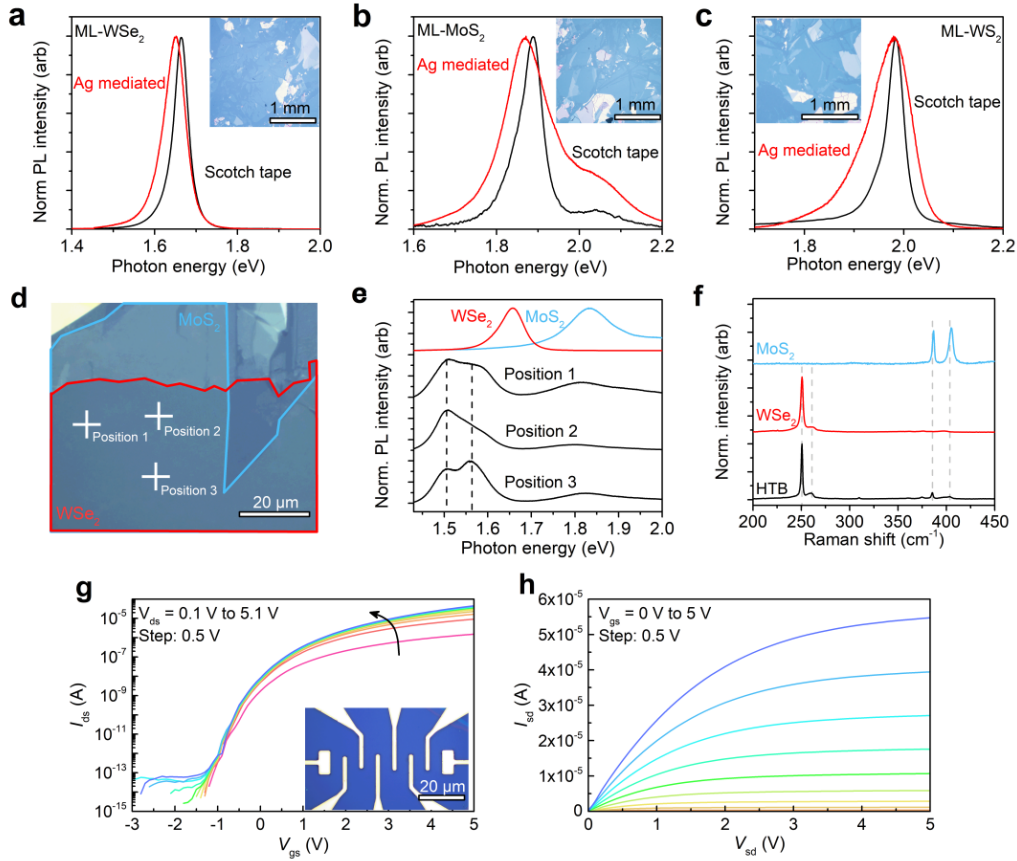

Fig. S12 Optical and electrical characterization of monolayer and heterobilayer samples transferred onto SiO$_2$/Si substrates. **a-c** PL spectra comparison for monolayer WSe$_2$, monolayer MoS$_2$, and monolayer WS$_2$ respectively, prepared via Ag-mediated exfoliation (red) and scotch tape method (black). **d** Optical microscope image of a MoS$_2$/WSe$_2$ heterobilayer. **e** Room temperature PL spectra of MoS$_2$/WSe$_2$ heterobilayer at 3 different sites of the overlap region and intralayer exciton emission at single layer regions. Two lower energy peaks emerged in the heterobilayer PL spectra that is unobservable in the single layer regions are considered IX emission. **f** Raman spectra of a MoS$_2$/WSe$_2$ heterobilayer acquired at the single layer and overlap regions in **d**, dashed gray lines are for visual clarity. **g** Transfer characteristic of a back-gated monolayer MoS$_2$ FET for V$_{ds}$ from 0.1 V to 5.1 V, step: 0.5 V, from which mobility ~15 cm$^2$ V$^{-1}$ s$^{-1}$ and on-off ratio ~10$^8$ are derived. **h** Output characteristic for the same device for V$_{gs}$ from 0 V to 5 V, step 0.5 V.

# Supplementary note 1: Raman evidence of CLQB between 2D crystals and Ag films

Theoretical results indicate that the interaction between many layered crystals and Ag atoms is stronger than the intrinsic crystal interlayer van der Waals interactions. This interaction between Ag and layered crystal is similar to the CLQB interaction between Au interface and layered crystals[21, 23]. For exfoliated 2D crystals on Ag film, in particular the bottom layers that have direct contact with Ag, the CLQB at the interface is expected to suppress or even extinct some of the low wavenumber (LW) interlayer Raman modes, resembling a pining-like effect as presented in our previous work[33]. For layered materials such as $MoS_2$ and $WSe_2$, their Raman activated interlayer modes are in-plane $E_{2g}^2$ shearing mode (C mode) and out-of-plane $B_{2g}^2$ intralayer breathing mode (LB mode).

Fig. S8a presents an optical microscope image of as-exfoliated $MoS_2$ 2D crystals on Ag film patterned with hole arrays, the layer number can be easily distinguished from the optical contrast, and also confirmed by intensity ratio between characteristic Raman peak ($A_{1g}$ and $E_{2g}^1$) and Si peak (~520.7 $cm^{-1}$) as shown in Fig. S8b. Fig. S8c shows LW Raman activated modes of suspended (left) and supported $MoS_2$ on Ag film (right), and the layer number of $MoS_2$ range from monolayer to 5L. The evolution of peak positions for C and LB modes are summarized in Fig. S8d. In comparison to previous reports[33-35], Raman shift of C mode and LB mode in suspended and supported $MoS_2$ show no major difference. However, there exhibits conspicuous suppression or extinction of multiple LW modes for $MoS_2$ on Ag film, especially for LB mode that cannot be distinguished even up to 4L, which supports the existence of CLQB at the $MoS_2$/Ag interface. Since the back-scattering setup also extinct or suppress LW Raman modes because substrates can hinder vibrations, samples directly exfoliated onto $SiO_2$/Si substrates were taken as reference, which can only weakly couple with 2D crystals via van der Waals interactions. As demonstrated in Fig. S8e, despite clear quenching in LW Raman modes, LB mode of 2L $MoS_2$ on $SiO_2$/Si substrate is still weakly distinguishable at ~40.03 cm$^{-1}$, which supports our speculation about the CLQBs induced pining-effect at the interface.

# Supplementary note 2: Transfer and characterization of exfoliated large-scale samples

Resembling the traditional wet transfer of 2D crystals, polymethyl methacrylate (PMMA) can be used as transfer media, and $NaI/I_2$ solution as a mild etchant to remove underlying Ag films without posing damage to the samples. However, organic residues will contaminate the sample surface, which is detrimental to establishing a good electrical contact or carrying out different types of surface characterizations. Here, we replace the Ti adhesive with a water-soluble polyvinylpyrrolidone (PVP) layer. Ag films carrying monolayer samples can detach from the sacrificial substrate in less than 10 mins if immersed into DI water. Detached Ag films are subsequently fished out by a glass slide and then cleaned for three times in DI water. The target substrates then pick up the Ag films in DI water, which is then heated up, cleaned, and etched by $NaI/I_2$ for approximately 6 hours to eliminate the Ag film. In order to be certain the optical properties of Ag-mediated samples are well preserved during the exfoliation and transfer procedures, monolayer $MoS_2$, $WS_2$ and $WSe_2$ are prepared via Ag-mediated exfoliation and traditional scotch tape method. Samples on Ag films are subsequently transferred onto another $SiO_2$/Si substrate for PL measurement. Fig. S12a-c shows the PL spectra of these monolayer samples on $SiO_2$/Si substrate prepared by Ag-mediated exfoliation (red) and scotch tape method (black), which demonstrate high quality of our Ag-mediated samples as the spectral shapes and peak positions are almost identical. The notable difference in emission peak width is likely resulted from the variable intensity ratio between neutral exciton, B exciton, and trion emission, which is reasonable given the fluctuations in the quality and defect densities of raw bulk crystals. Insets of Fig. S12a-c are optical microscope images of transferred samples, which remains intact and their sizes well preserved, making them available for multiple purposes such as stacking of type II heterobilayers (HTBs) for investigation of interlayer exciton (IX) emission or the fabrication of nano devices such as FETs.

For manual stacking of HTBs, if the interlayer twisted angle is not specifically demanded, alignment procedures under a transfer station is unnecessary due to extraordinarily large area of our transferred samples. Following the same process, another detached and cleaned Ag film containing 2D crystals just needed to be successively fished out by the previous substrate. By this method, monolayer $MoS_2$/monolayer $WSe_2$ type II HTBs are prepared, and subsequently PL and Raman spectroscopies are conducted. As shown in Fig. S12e, the room temperature PL spectra of $MoS_2$/$WSe_2$ HTB contains three major peaks, the two lower energy peaks around 1.50 and 1.56 eV can be considered as IX emission peaks. These peaks corresponds to the transition between electrons in K valleys of $MoS_2$ and holes resulted from the hybridized states at $\Gamma$ points, which is not detectable in either monolayer $MoS_2$ or monolayer $WSe_2$. The Raman spectrum of HTB is also observable for characteristic intralayer vibrational modes of both monolayer films. These results demonstrate that fine interface quality and interlayer coupling has been perfectly established in our manually stacked HTBs, enabling potential investigations of optical and vibrational properties in manually stacked twisted or well aligned heterostructures and homostructures.

For the purpose of demonstrating device application potentials of our exfoliated and transferred samples, we fabricated a back-gated FET device based on the transferred monolayer $MoS_2$. Standard photolithography process, followed by thermal evaporation and liftoff were employed to define the 50 nm Au/5 nm Ti electrodes (inset of Fig. S12g). Transfer and output characteristic curves are shown in Fig. S12g and S12h, respectively, from which large on-off ratio ~$10^8$ and carrier mobility ~20 $cm^2$ $V^{-1}$ $s^{-1}$ can be derived, which is comparable to most monolayer $MoS_2$ back-gated devices. The electrical transport measurements indicate that the electrical properties are well preserved during the transfer procedures.

# Supplementary note 3: PVP functionalized transfer of as-exfoliated samples and heterobilayer stacking

A wafer-scale SiO$_2$/Si substrate was first spin-coated with PVP solution (PVP powder Mol. wt. 38000 10% wt. in 1:1 ethanol/acetonitrile solution, 3000 rpm 30s, acceleration 2000 rpm/s twice and baked at 150 °C for 5 min). 150 nm Ag film was then deposited in an in-glove-box thermal evaporation system at 1 Å s$^{-1}$ rate. After the sample exfoliation, the substrates carrying Ag films and samples are placed upside down and then immersed into DI water to dissolve the PVP layer. After roughly 10 min, the Ag films with samples will be detached from the substrate and then fished out by glass slides, which is subsequently cleaned for three times with DI water. After that the target substrates were used to fish out the Ag films and is then heated on a hotplate at 60 °C to evaporate the liquid underneath. Afterwards, substrates with Ag film were cleaned by oxygen plasma for 2 min to remove possible PVP residues, and immersed into I$_2$/NaI solution (2.5 g I$_2$ and 10 g NaI in 100ml DI water) for approximately 6 hours to remove the Ag films. Substrates carrying large-scale monolayers are then cleaned successively by acetone, isopropanol, and DI water to eliminate potential ion residues.

Stacking of heterobilayers started with the transfer of as-exfoliated samples by the same PVP functionalized transfer method illustrated above. Monolayer samples were successively transferred onto a new SiO$_2$/Si substrate, each time followed by a lift-off and annealing process to remove interfacial organic residues and to ensure that good interlayer adhesion and coupling were established. Since our samples were macroscopic in size, no specific alignment procedure was needed, as long as interlayer twisted angle does not need to be specified.

# Supplementary Table 1. Calculated energies of 18 considered 2D materials

| 2D materials | Ads. Energy on Ag(111) (eV per unit cell) | Interlayer Coupling Energy (eV per unit cell) | Ads. Energy on Ag(111) (eV Å-2) | Interlayer Coupling Energy (eV Å-2) | $R_{MA/IL}$ | Magnetic Structure* |
|---|---|---|---|---|---|---|
| Graphene | 0.134 | 0.118 | 0.0255 | 0.0225 | 1.13 | NM |
| h-BN | 0.134 | 0.136 | 0.0246 | 0.0250 | 0.99 | NM |
| P(Black) | 0.840 | 0.484 | 0.0583 | 0.0334 | 1.74 | NM |
| $CrCl_3$ | 0.854 | 0.550 | 0.0279 | 0.0180 | 1.55 | FM |
| $RuCl_3$ | 0.776 | 0.574 | 0.0461 | 0.0368 | 1.35 | NM |
| $MoS_2$ | 0.389 | 0.229 | 0.0448 | 0.0263 | 1.70 | NM |
| $MoSe_2$ | 0.426 | 0.243 | 0.0453 | 0.0258 | 1.76 | NM |
| $MoTe_2$ | 0.509 | 0.281 | 0.0474 | 0.0261 | 1.82 | NM |
| FeSe | 0.515 | 0.381 | 0.0391 | 0.0285 | 1.37 | NM |
| $SnSe_2$ | 0.795 | 0.275 | 0.0625 | 0.0214 | 2.92 | NM |
| $SnS_2$ | 0.641 | 0.243 | 0.0550 | 0.0207 | 2.66 | NM |
| $WS_2$ | 0.350 | 0.227 | 0.0395 | 0.0261 | 1.51 | NM |
| $WTe_2$ | 1.033 | 0.516 | 0.0474 | 0.0235 | 2.02 | NM |
| $MnBi_2Te_4$ | 0.812 | 0.368 | 0.0510 | 0.0229 | 2.23 | FM |
| $WSe_2$ | 0.385 | 0.241 | 0.0409 | 0.0256 | 1.60 | NM |
| 2H-$TaS_2$ | 0.683 | 0.249 | 0.0716 | 0.0262 | 2.73 | NM |
| 1T-$TaS_2$ | 0.528 | 0.278 | 0.0543 | 0.0286 | 1.90 | NM |
| Fe$_3$GeTe$_2$ | 0.739 | 0.434 | 0.0556 | 0.0327 | 1.70 | FM |

*NM: Non-magnetic; FM: Ferromagnetic